\documentclass[preprint2]{aastex}

\usepackage{amsmath}
\usepackage{txfonts}

\usepackage{mdwlist}

\makecompactlist{thebibliography*}{shortbib}

\makeatletter
\newcommand\pss{\ref@jnl{Planet. and Space Sci.}}
\newcommand\pr{\ref@jnl{Physics Reports}}
\newcommand\pl{\ref@jnl{Physics Letters}}
\newcommand\asr{\ref@jnl{Adv. Space Res.}}
\newcommand\ass{\ref@jnl{Astrophys. Space Sci.}}
\newcommand\aea{\ref@jnl{A\&A}}
\makeatother

\newcommand{\um}[1]{\ensuremath{\, \mathrm{#1}}}
\newcommand{\de}{\ensuremath{\, \mathrm{d}}}

\begin{document}

\title{The Local Interstellar Spectrum of Cosmic Ray Electrons}

\author{Diego Casadei\altaffilmark{1} and Veronica Bindi}

\affil{INFN, Sezione di Bologna,%\\
      Via Irnerio 46, I-40126 Bologna, Italy}

\altaffiltext{1}{\url{Diego.Casadei@bo.infn.it}}

\shorttitle{The LIS of CR Electrons}

\shortauthors{Casadei, Bindi}

\date{\centering Feb 9, 2004}

\begin{abstract}
 The direct measurements of electrons and positrons over the last 30
 years, corrected for the solar effect in the force-field
 approximation, are considered.  The resulting overall electron
 spectrum may be fitted with a single power law above few GeV with
 spectral index ($\gamma_{-} = 3.41 \pm 0.02$), consistent with the
 spectral index of the positron spectrum ($\gamma_{+} = 3.40 \pm
 0.06$), therefore suggesting a common acceleration process for both
 species.  We propose that the engine was a shock wave originating
 from the last supernova explosion among those that formed the local
 bubble.  In addition, at low energy, the electron spectrum measured
 during the last $A+$ solar phase is damped, whereas the positron
 spectrum is well represented by a single power law down to the lowest
 inferred local interstellar energy (0.8 GeV).  We suggest that this
 difference arises from a time- and charge-dependent effect of the
 solar modulation that is not taken into account by the force-field
 approximation. 
\newline
 \emph{OBSOLETE: please refer to ApJ 612 (2004) 262-267, that is the
 final version of this work.}
\end{abstract}


\section{INTRODUCTION}

 Cosmic ray (CR) electrons are probably accelerated by the same
 engines that accelerate CR protons and nuclei (galactic supernova
 explosions), whereas positrons are believed to be produced by the
 interactions of CR with the interstellar medium (ISM)
 \citep{berezinskii90}.  However, electrons and positrons differ
 significantly from hadrons for what concerns the energy lost during
 the propagation through the ISM: in addition to ionization losses in
 the ISM, due to their small mass they suffer large radiative losses
 due to electromagnetic processes as synchrotron radiation, inverse
 Compton scattering, and bremsstrahlung.  In general, all these
 effects induce an energy loss rate that increases with the energy as:
\begin{equation}
  - \left( \frac{\de E}{\de t} \right)_\text{tot} = a E^2 + b E \; ,
\end{equation}
 whereas radiative losses are usually not important for CR protons and
 nuclei.  Therefore, CR electrons can not diffuse for large distances:
 the radiative losses limit their range to \citep{kobayashi01}:
\begin{equation}
  r \sim 1 \um{kpc} \, \left( \frac{E}{1 \um{TeV}} \right)^{-1} \; .
\end{equation}
 In addition, the source must be relatively recent in order not to have
 important radiative losses: 
\begin{equation}
  t_{\mathrm{rad}} \approx 2.1 \times 10^5 \left(
    \frac{E}{1 \um{TeV}} \right)^{-1} \um{years.}
\end{equation}
 On the contrary, CR protons have a long residence time in the Galaxy
 ($\sim 10^7$ years) and sample a large fraction ($\approx 1/3$) of
 the disk and halo \citep{maurin02}.

\begin{figure}[t!]
\plotone{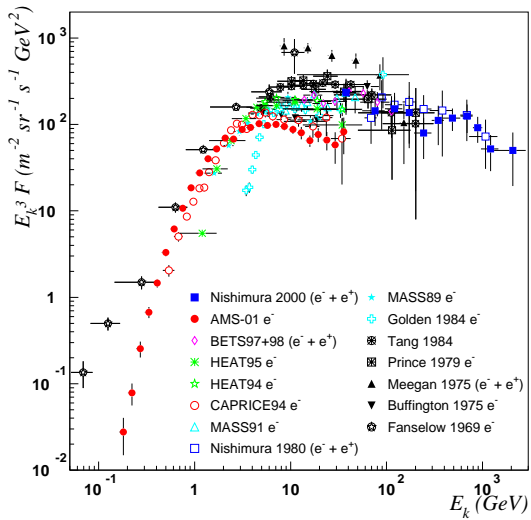}
\caption{Direct measurements of the CR electrons. \label{allx3}}
\end{figure}

 Many authors emphasized that the CR electron spectrum can not be
 considered representative of a galactic average, but it has to
 reflect the recent history of the solar system neighborhood (see for
 example \citealt{torii00,muller01,stephens01}), expecially in the
 highest energy range.  However, the existing data on direct CR
 electron and positron measurements (figure~\ref{allx3},
 table~\ref{tbl-1}) have been interpreted in a number of different
 ways.  \citet{atoyan95} considered the overall electron spectrum as
 the sum of a galactic average and a single source component, and
 \citet{kobayashi01} concluded that Vela (distance 250 pc, age 20 ky)
 should have been produced this component.  However, \citet{guetta03}
 stressed that the high energy photons coming from Vela are likely due
 to hadronic processes instead of inverse Compton scattering.  Another
 possible source could be the Monogem pulsar \citep{thorsett03}, whose
 distance and age match the prediction of the model by
 \citet{erlykin97}, who attributed the knee feature to a single nearby
 source.  On the other hand, \citet{pohl98} considered the measured
 spectrum as the superposition of many ``single-source'' components,
 producing several spectral breaks and a time dependent electron
 spectrum.  Finally, we recall the analysis made by \citet{muller01},
 who considered different behaviors in different energy ranges,
 depending on the energy-dependent diffusion coefficient, the energy
 losses and the distance to the nearest sources.

 In this paper we consider only the recent measurements of the CR
 electron and positron spectra (starting with MASS89) and the highest
 energy data from \citet{nishimura80,nishimura00}.

 We corrected the measured spectra for the solar modulation effect in
 the simplest framework: the spherical symmetric force-field
 approximation (section~\ref{sol-mod}).  Among these experiments, only
 CAPRICE94 and AMS-01 were able to measure both protons and electrons
 (and positrons).  In this case we use the proton spectrum to measure
 the solar modulation parameter, whereas neutron monitor rates are
 used to demodulate the results of the other detectors.  Even after
 this correction, there is still a large spread between measurements
 of the electron spectrum in the 10--100 GeV range, where solar
 effects are very small.  Following \citet{muller01} and
 \citet{duvernois01}, we rescaled all normalization constants in order
 to reduce this spread, in the hypothesis that systematic errors do
 not affect sensibly the measured spectral indices.

 The resulting spectrum has been fitted with a single power law from 3
 GeV up to 2 TeV, and with a single source model above 10 GeV
 (section~\ref{results}).  On the other hand, we fitted the positron
 spectrum with a single power law in kinetic energy (without any
 renormalization), obtaining the same spectral index as the power law
 fit of the electrons.  In section~\ref{discussion} we speculate about
 the possibility that both electrons and positrons have been
 accelerated by the same engine.  In addition, we considered the
 measured e$^+$/e$^-$ ratio at low energy as the effect of the solar
 modulation with a finite propagation time.

\section{SOLAR MODULATION}\label{sol-mod}

 In order to estimate the local interstellar spectrum (LIS), it is
 necessary to choose a model for the solar modulation of cosmic rays.
 We used the spherically symmetric adiabatic model of \citet{parker65}
 and \citet{gleeson67,gleeson68}, in which the charged particles have
 a diffusion coefficient that depends on their rigidity $R$ and on the
 distance $r$ from the Sun, with a typical time scale of 11 years
 given by the solar activity half-cycle.  Over this time scale, the
 variation of the incoming cosmic ray flux from the Galaxy is
 negligible, and the problem reduces to the one-dimensional diffusion
 of charged particles that are adiabatically decelerated by the solar
 wind.  The analytical solution is possible \citep{gleeson68}, and the
 measured differential flux $J(r,E,t)$ of particles with total energy
 $E$ and mass $E_0/c^2$ is:
\begin{equation}\label{eq-sol-mod}
 J(r,E,t) = 
     \frac{E^2 - E_0^2}{(\Phi(t)+E)^2 - E_0^2} \,
     J(\infty,\Phi(t)+E)
\end{equation}
 where $J(\infty,\Phi(t)+E)$ is the stationary flux outside the
 Eliosphere of particles with energy $E' = \Phi(t) + E$, and $\Phi(t)$
 is the energy lost by the particles during their travel. 

 In this ``force-field'' model, one can write $\Phi(t) = |Z| e
 \phi(t)$ where $|Z|e$ is the absolute value of the particle charge
 and the solar modulation parameter $\phi(t)$ can be expressed as
 function of the diffusion coefficient and the solar wind velocity.
 Actually, the full force-field parameter is the adimensional term
 $\phi/[\beta\,\kappa_2(R)]$, where $\kappa_2(R)$ is the
 rigidity-dependent term of the diffusion coefficient
 \citep{gleeson73}.  However, \citet{caballero03} emphasized that
 above 200 MeV $\kappa_2 \propto R$ and $\beta \approx 1$, hence using
 $\phi$ (interpreted as the rigidity loss of incoming particles) as
 the single parameter is a good approximation to the numerical
 solution of the spherically-symmetric transport equation that
 includes adiabatic energy losses correctly.  In this paper, we
 considered $\phi$ a free parameter to be measured independently

\begin{figure}[t!]
\plotone{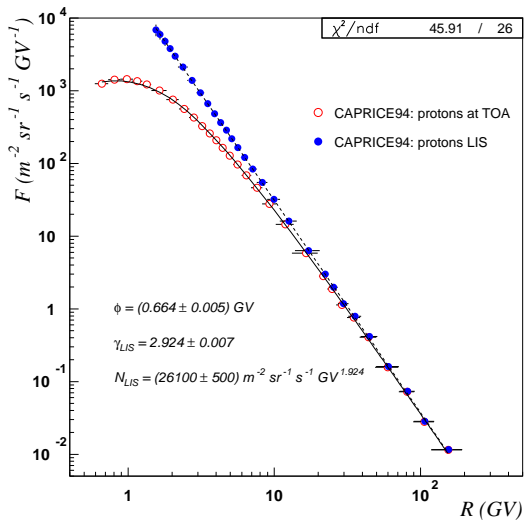}
\caption{Proton spectrum measured by CAPRICE94 \citep{boezio99} and
  inferred LIS. \label{cap94pr}} 
\end{figure}

\begin{figure}[t!]
\plotone{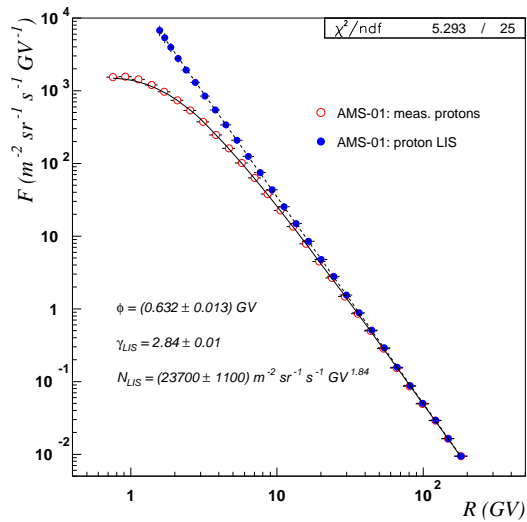}
\caption{Proton spectrum measured by AMS-01 \citep{ams00pr2} and
  inferred LIS. \label{ams98pr}}
\end{figure}

 A way to quantify the effects of the solar activity on the primary
 cosmic ray fluxes is to use neutron counters placed in high
 mountains, that are able to measure the rate of interactions of
 primary particles with the Earth atmosphere by detecting the produced
 secondary neutrons.  The neutron rate is anti-correlated to the
 number of sun-spots (that can be considered an index of solar
 activity) while it is correlated to the interplanetary cosmic ray
 flux \citep{clem96}.  However, the best way to find the value of the
 solar modulation parameter is to use the proton flux measured at the
 top of the Earth atmosphere \citep{boezio99}.  This method has been
 adopted only with the detectors that were able to measure both
 protons and electrons at the same time: CAPRICE94 ($\phi = 664\pm5$
 MV, figure~\ref{cap94pr}) and AMS-01 ($\phi = 632 \pm 13$ MV,
 figure~\ref{ams98pr}).  The value of $\phi$ found with protons or
 neutron rates is then used to correct the electron and positron
 spectra, neglecting charge-dependent modulation effects.

 The force-field model is a strong simplification of the propagation
 process inside the eliosphere (see for example \citealt{clem00}) and
 must be considered only a first approximation working well only for
 high enough particle rigidity and for the inner eliosphere
 \citep{caballero03}.  However, the very low value of the chi-square
 compared to the number of degrees of freedom of the AMS-01 proton fit
 with a modulated single power-law is significant\footnote{The fit is
 impressive also with CR protons measured by BESS in 1998, 1999, and
 2000.}.  There is absolutely no indication that the cosmic ray
 protons spectrum outside the eliosphere is not a single power law in
 rigidity below 200 GV, an important constraint for propagation
 models.  In any case, we use the force-field model only as a way to
 compare different measurements, without stating anything about the
 underlying physical processes of electron diffusion.

 \citet{ams00pr2} obtained the primary cosmic ray proton spectrum by
 fitting the measured AMS-01 spectrum above 10 GV with a single power
 law with unknown spectral index and normalization, assuming that the
 solar modulation effects can be neglected above this rigidity: their
 result is $\gamma = 2.79 \pm 0.02$ (summing in quadrature the quoted
 statistical and systematic errors).  However we obtain a slightly
 different spectral index ($\gamma = 2.84 \pm 0.01$) by fitting with a
 function of three parameters that takes $\phi$ explicitly into
 account.  This method is more reliable and demonstrates that when the
 experimental uncertainties are as low as those of AMS-01 the effects
 of the solar activity can be appreciable up to few tens of GV.

\begin{figure}[t!]
\plotone{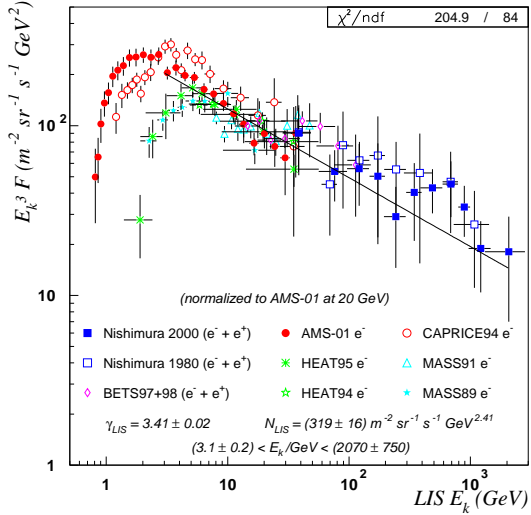}
\caption{Inferred LIS of CR electrons.  The measurements have been
  renormalized to AMS-01 at 20 GeV.  The data points are fitted by a
  single power law in kinetic energy from about 3 GeV to 2
  TeV. \label{reclisnormx3}}
\end{figure}

\section{RESULTS}\label{results}

 We do not make use of the eldest measurements, apart from data
 obtained by \citet{nishimura80}, that agree with the recent update by
 \citet{nishimura00} and are almost insensitive to the solar
 modulation.  Notwithstanding, the existing data sets have a non
 negligible spread in the 10--100 GeV range, where the measured fluxes
 at the same energy may differ even by a factor of 10.  However,
 supposing that the main source of uncertainty comes from the
 correction for the residual atmosphere and that this systematic
 effect is not greatly dependent on the spectral index, one can adjust
 the normalization constants of the various experiments to reduce the
 spread \citep{duvernois01,muller01}.  Incidentally, this
 automatically adjusts the systematic shift due to the positron
 component for the detectors that were not able to separate positive
 from negative charges (see table~\ref{tbl-1}), in the hypothesis that
 the positron fraction reaches a constant value at high energies.

\begin{figure}[t!]
\plotone{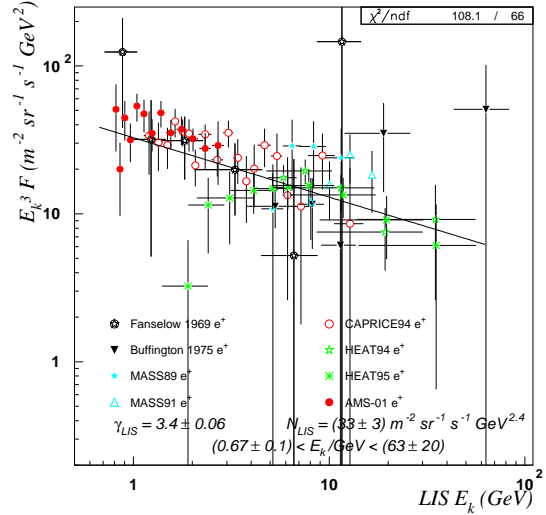}
\caption{Inferred LIS of CR positrons.  The data points are fitted by
  a single power law in kinetic energy from 0.7 GeV to about 60
  GeV, without any renormalization. \label{plisallx3}}
\end{figure}

 The resulting local interstellar electron spectrum is shown in
 figure~\ref{reclisnormx3}, where the inferred local interstellar
 spectrum of the electrons has been fitted with a single power law
 from 3 GeV to 2 TeV.  On the other hand, all positron measurements
 agree well without renormalization, and are well represented by a
 single power law in kinetic energy (figure~\ref{plisallx3}), the only
 exception being the low energy data points obtained by HEAT95, whose
 electron measurement at low energy is also in disagreement with other
 experiments.

 As shown in figures \ref{reclisnormx3} and \ref{plisallx3}, the LIS
 inferred from the HEAT95 measurement \citep{duvernois01} differ
 significantly from the other experiments below few GeV, both for
 electrons and positrons.  The electron measurement done by MASS89
 \citep{golden94} is the only one that agrees with HEAT95 at low
 energy, even though the latter took data during a solar minimum
 whereas the former operated during a solar maximum.  Hence it seems
 unlikely that the discrepancy between these two experiments and the
 other ones is caused by a bad demodulation procedure.  Rather, it
 seems that unknown systematics affect these results.

\subsection{The local interstellar spectrum}

 Both the electron and the positron spectra can be fitted with single
 power-laws, and the two spectral indices obtained by the best fit
 procedures are compatible: the local interstellar spectra have
 $\gamma_{-} = 3.41 \pm 0.02$ for electrons above 3 GeV and
 $\gamma_{+} = 3.40 \pm 0.06$ for positrons up to 60 GeV.  The single
 power-law fit of the electron LIS has a not very good chi-square
 compared to the number of degrees of freedom.  However, with
 exception of CAPRICE94 and AMS-01, that quoted both statistical and
 systematic errors (here summed in quadrature), it must be noted that
 all experiments quoted only statistical errors in the numerical
 tables.  The systematics are usually of the order of 10\% at least,
 and were not considered in the fit.  Hence the obtained chi-square
 should not be considered that bad.  On the other hand, the chi-square
 of the positron LIS fit with a single power-law is quite good, taking
 into consideration statistical errors only (apart from CAPRICE94 and
 AMS-01).

\begin{figure}[t!]
\plotone{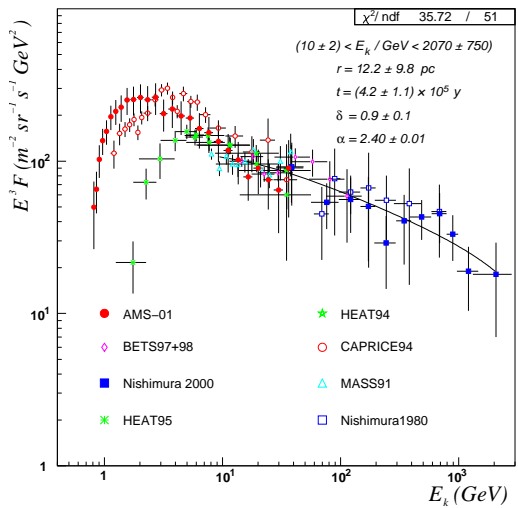}
\caption{Inferred LIS of CR electrons fitted with the single source
  model by \citet{atoyan95}. \label{atoyanx3}}
\end{figure}

 The highest energy data points \citep{nishimura80,nishimura00} come
 all from the same kind of detector, using the emulsion chambre
 technique.  Several authors have seen a cut-off at about 700--800 GeV
 in the first measurement by \citet{nishimura80}.  However, the last
 published data by \citet{nishimura00} extend the energy range up to 2
 TeV, making it very difficult to affirm that a cut-off does indeed
 exist below 1 TeV.  In order to test the cut-off hypothesis, we
 obtained the best fit of the electron LIS using the single source
 model by \citet{atoyan95} (figure~\ref{atoyanx3}).  This quite
 complex function depends on the source distance $r$ and age $t$, on
 the injection spectrum spectral index $\alpha$, and on the exponent
 $\delta$ of the diffusion coefficient $D(E) = D_0 (1 +
 E/E_0)^\delta$, and should well represent data at high energy, where
 the single source is expected to dominate over the galactic component
 (important at lower energies).  In our knowledge, this paper is the
 first one where this function is used in a best fit procedure: other
 authors simply computed it for nearby source candidates and compared
 the result with the measurement.

\begin{figure}[t!]
\plotone{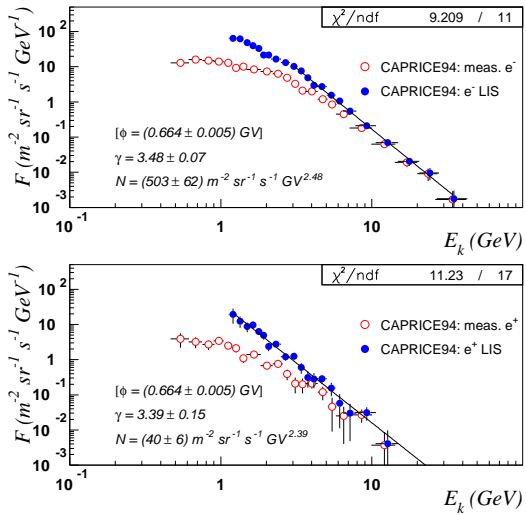}
\caption{Measured and local interstellar flux of CAPRICE94 electrons
(upper panel) and positrons (lower panel). \label{caprice94lis}}
\end{figure}

 We used the single source spectrum to fit the electron LIS above 10
 GeV, obtaining a very good chi-square value.  The results is that the
 single source must be recent [$t = (4.2 \pm 1.1) \times 10^5$ y] and
 very near to the solar system ($r = 12.2 \pm 9.8$ pc).  In addition,
 the source spectral index is 2.4 and the exponent of the diffusion
 coefficient appears to be $\delta = 0.9 \pm 0.1$, greater than the
 usual adopted value of 0.6 but consistent with recent simulations by
 \citet{maurin02}.  However, in case of a single source dominating the
 LIS at high energy over the galactic component, we should see a
 spectral change at 10--20 GeV, that is not seen by CAPRICE94
 (figure~\ref{caprice94lis}) and AMS-01 (figure~\ref{ams98elpo}).

\subsection{The low energy spectrum}

\begin{figure}[t!]
\plotone{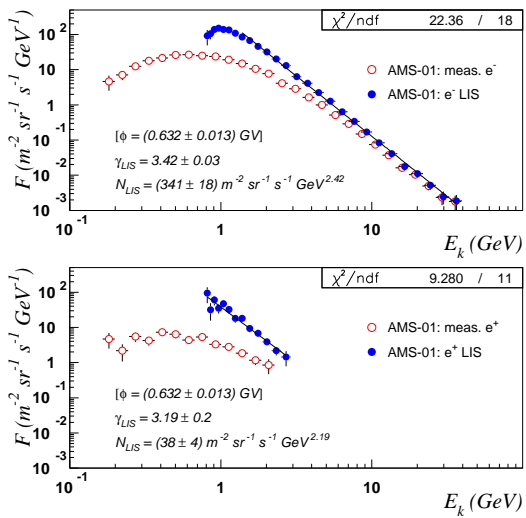}
\caption{Measured and local interstellar flux of AMS-01
electrons (upper panel) and positrons (lower panel). \label{ams98elpo}}
\end{figure}

 Using the same value for $\phi$ as given by the protons fit and
 neglecting any variation in the rigidity dependence of the electron
 diffusion coefficient, it is possible to correct for the solar
 effects the electron and positron primary spectra measured by
 CAPRICE94 \citep{boezio00} and AMS-01 \citep{ams00el} using equation
 (\ref{eq-sol-mod}), as shown in figures \ref{caprice94lis} and
 \ref{ams98elpo}.  We would expect this approximation to break down at
 low energies, due to the change of the diffusion coefficient of e$^+$
 and e$^-$ below few GV \citep{potgieter99}.  However, the damping at
 low energies in the local interstellar spectrum of electrons is not
 visible in the positron LIS, similar to the proton LIS shown in
 figures \ref{cap94pr} and \ref{ams98pr}.  Instead, the force-field
 approximation is working very well both for protons and positrons
 over the whole energy range, and for electrons above few GeV.

 The next order approximation should take into account the particle
 charge \citep{bieber99}.  This effect is expected to be more evident
 at lower energies and could explain the feature visible in the
 electron LIS but not in the positrons LIS.  In addition, the electron
 LIS inferred from CAPRICE94 data is a power-law above 3 GeV, whereas
 the AMS-01 spectrum is damped below 2 GeV.  In 1994 the CR flux was
 recovering from the minimum occurred in 1991, while in 1998 the
 situation was opposite: the flux was decreasing from the maximum in
 1997 towards the following minimum (in 2000).  Because the solar
 modulation has a necessary delay connected with the propagation time
 of the magnetic irregularities from the Sun to the eliopause, it is
 evident that the damping should be higher just following a CR minimum
 than for periods preceding a minimum.  This can explain why AMS-01
 data show a departure from the single power law at lower energies
 than CAPRICE94.  However, this effect is not observed in the proton
 and positron spectra, hence it has to be connected with the particle
 charge sign. 

 During the nineties, the Sun was in a positive half-cycle, when the
 magnetic field emerging from the North Pole of the Sun pointed
 outward \citep{bieber99}.  In this case, the antisymmetry of the
 Parker field above and below the equatorial plane produces drift
 velocity fields for positive particles that converge on the equator,
 whereas negative particles diverge from it \citep{clem00}.  During
 the negative polarity solar half-cycle started in 2001 the situation
 is reversed, and the damping at low energy should be visible with
 positrons but not in the electron spectra.  Measurements from future
 experiments will allow to check this behavior.

\begin{figure}[t!]
\plotone{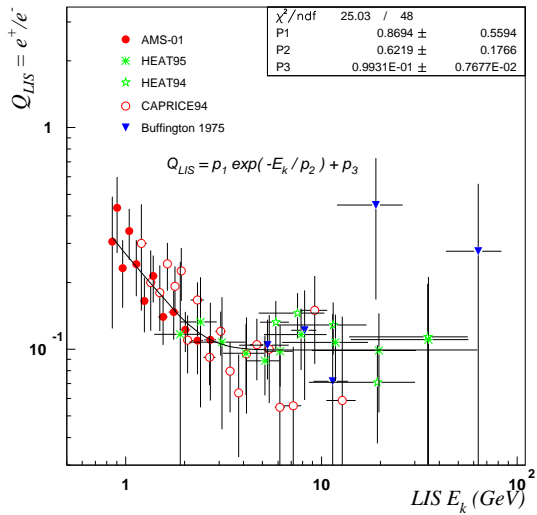}
\caption{Positrons to electrons ratio as function of the local
  interstellar energy. \label{qrlisnormrec}}
\end{figure}

 The positive half solar cycle positron to electron ratio $Q$ is shown
 in figure~\ref{qrlisnormrec}, where we computed it starting from the
 absolute fluxes (we did not include the results of experiments that
 did not published absolute fluxes).  The ratio exponentially
 decreases from about 1:3 to 1:10 with increasing local interstellar
 energy from 0.7 GeV to about 4 GeV (the most interesting region for
 the study of the diffusion coefficient), then it seems to be flat.
 The possible ``bump'' seen by \citet{buffington75} has not been
 confirmed by any other experiment: the ratio flattens at high
 energies.

\section{DISCUSSION}\label{discussion}

 The cosmic ray electron flux is being measured since 30 years with
 balloon and satellite experiments over a wide energy range.  The
 results from different detectors are not always compatible within
 quoted errors, expecially for the eldest experiments.  In addition
 the energy range below 100 GeV is covered by many experiments, but
 above this level only data from \citet{nishimura80} and
 \citet{nishimura00} are available up to now.  They were obtained
 using the emulsion chambre technique, but in the near future
 different methods will be adopted.  For example the PAMELA
 \citep{pamela} and AMS-02 \citep{ams02} detectors will be magnetic
 spectrometers with proton-electron discrimination obtained using an
 electromagnetic calorimeter and a transition radiation detector.

 After renormalizing the different measurements at 20 GeV, the single
 power-law fit of the LIS of CR electrons gives a spectral index of
 $\gamma_{-} = 3.41 \pm 0.02$, from 3 GeV up to 2 TeV.  On the other
 hand, the positron LIS can be fitted by a single power-law over the
 whole energy range covered by existing measurements, without any
 renormalization.  The positron spectral index $\gamma_{+} = 3.40 \pm
 0.06$ is remarkably equal to the electron value within the small
 errors.  No evident cut-off can be found below 1--2 TeV in the
 electron spectrum, indicating that their source must be recent (at
 most few $10^5$ years) and nearby (few 100 parsecs).  An intriguing
 possibility is that both electrons and positrons were accelerated by
 the same source, according to the following scenario.

 Electrons are produced by supernova explosions and by the
 interactions of CR hadrons (mostly protons and He nuclei) with the
 ISM, whereas positrons are mostly secondary products of these
 interactions.  The spectral index of protons is about 2.8, and this
 is also the injection spectral index of positrons.  In stationary
 conditions, the positrons loose energy by radiative processes and
 their spectral index is lowered by one, becoming roughly equal to
 3.8.  This average galactic population was present also in the solar
 system neighborhood, where they were reaccelerated by the recent
 shock wave that accelerated electrons, likely produced by the last
 supernova among those that formed the local bubble
 \citep{erlykin97,maiz01,benitez02}.  The ``natural'' spectral index
 produced by this engine has to be $\alpha \lesssim 3.4$, equal to or
 slightly harder than the measured spectral index of positrons and
 electrons, because the acceleration is recent enough for the spectrum
 not to have been distorted significantly by radiative losses.  The
 injection spectral index is also harder than the pre-existing
 positron spectral index, thus it is possible that positrons gained
 energy during the reacceleration \citep{voelk03}.

 The shock wave compression ratio was $R = (\alpha+2)/(\alpha-1)
 \gtrsim 2.25$, i.e.\ it was ``softer'' than the usual compression
 ratio $R=4$ used in the literature \citep{schli03}, that corresponds
 to a source spectral index $\alpha =2$ and to a synchrotron spectral
 index $a = (\alpha -1)/2 = 0.5$.  This value for $a$ is the average
 value of the Green's catalog of galactic supernova remnants
 \citep{green01}, whereas the value inferred from the measured
 positron and electron spectra ($a \lesssim 1.2$) is higher than all
 measurements reported in this catalog.  Hence, we do not expect to
 see important effects on the synchrotron background due to the shock
 wave that accelerated electrons and positrons.  In addition, the
 proton spectrum should not be affected at all, for two reasons:
 first, protons have a long residence time in the Galaxy and sample
 about one third of its disk.  Hence one can estimate the number $N$
 of supernovae that contributed to the proton spectrum considering the
 rate of supernova explosions in one third of the galactic disk as
 $\tau^{-1} \approx (100 \,\mathrm{y})^{-1}$ and the residence time as
 $T \approx 10^7$~y, then $N = T \tau^{-1} \approx 10^5$.  Thus a
 single source will tipically contribute a very small fraction of the
 CR protons.  Second, the ``natural'' spectral index produced by our
 source is higher than the proton index, i.e.\ its spectrum is softer.
 In this case, one does not expect such a weak reacceleration to
 increase the existing spectral index: it remains the same as before
 the shock \citep{voelk03}.  A source with no sensible effects on the
 hadronic component of galactic cosmic rays is also compatible with
 the absence of a persistent negative trend in the CR intensity,
 confirmed by the analysis of \citet{usoskin03}, who considered the CR
 flux over the last 400 years.

 At low energy, the extrapolated local interstellar spectrum of CR
 electrons is flattened, whereas the positron spectrum given by
 CAPRICE94 and AMS-01 is still well represented by a single power law
 down to the lowest inferred LIS energy (0.7 GeV).  In addition, the
 damping appeared at 3 GeV in 1994, following a solar maximum, and at
 2 GeV in 1998, just before the solar minimum.  A charge- and
 time-dependent solar modulation effect
 \citep{clem96,bieber99,heber00,potgieter01,webber01,clem03} with
 finite propagation velocity could explain this difference, though new
 measurements with high accuracy in the negative polarity solar
 half-cycle (where the behavior of electrons and positrons should be
 exchanged) are needed to confirm it, and several measurements in
 different times or long duration missions are needed to quantify this
 first order effect.  Future experiments PAMELA and AMS-02 will be
 able to measure the absolute fluxes of positrons and electrons up to
 few hundreds GeV during several years, and will be decisive to
 understand this phenomenon.

\acknowledgments

 The authors wish to thank the AMS Collaboraion for having provided
 the numerical tables of the e$^+$ and e$^-$ spectra, and S. Cecchini,
 V.A. Dogiel, F. Donato, R. Fanti, L. Foschini, D. M\"uller, F. Palmonari,
 V. Vitale, and H.J. V\"olk for the useful discussions about this
 work.

%\begin{thebibliography}{}
\begin{thebibliography*}

\bibitem[Alcaraz et al.(2000a)]{ams00pr1}
Alcaraz, J., et al. 2000a, \pl, B472, 215

\bibitem[Alcaraz et al.(2000b)]{ams00el}
Alcaraz, J., et al. 2000b, \pl, B484, 10

\bibitem[Alcaraz et al.(2000c)]{ams00pr2}
Alcaraz, J., et al. 2000c, \pl, B490, 27

\bibitem[Aguillar et al.(2002)]{ams02}
Aguilar, M., et al. 2002, \pr vol. 366, no. 6, 331

\bibitem[Atoyan et al.(1995)]{atoyan95}
Atoyan, A.M., et al.\ 1995, Phys.\ Rev.\ D52, 6, 3256

\bibitem[Ben\'{\i}tez et al.(2002)]{benitez02}
Ben\'{\i}tez, N. et al.\ 2002, Phys. Rev. Lett. 88, 8, 081101

\bibitem[Berezinskii et al.(1990)]{berezinskii90}
Berezinskii, V.S., et al.\ 1990,
\emph{Astrophysics of Cosmic Rays},
Amsterdam, North-Holland Publ.\ Co.

\bibitem[Bieber et al.(1999)]{bieber99}
Bieber, J.W., et al. 1999, \prl, vol.\ 83, no.\ 4, 674

\bibitem[Boezio et al.(1999)]{boezio99}
Boezio, M., et al. 1999, \apj, 518, 457

\bibitem[Boezio et al.(2000)]{boezio00}
Boezio, M., et al. 2000, \apj, 532, 653

\bibitem[Bonvicini et al.(2001)]{pamela}
Bonvicini, V., et al. 2001, NIM A 461, 262

\bibitem[Buffington et al.(1975)]{buffington75}
Buffington, A., Orth, C. D., \& Smoot, G. F. 1975, \apj, 199, 669 

\bibitem[Caballero-Lopez \& Moraal(2003)]{caballero03}
Caballero-Lopez, R.A., \& Moraal, H. 2003, subm.\ to JGR

\bibitem[Clem et al.(1975)]{clem96}
Clem, J.M., et al. 1996, \apj, 464, 507

\bibitem[Clem et al.(2000)]{clem00}
Clem, J.M., et al. 2000, \jgr, vol.\ 105, no.\ A10, 23099

\bibitem[Clem \& Evenson(2003)]{clem03}
Clem, J.M.  \& Evenson, P. 2003, Proc.\ of the 28$^\mathrm{th}$ ICRC,
SH 3.4, 4023

\bibitem[Du Vernois et al.(2001)]{duvernois01}
Du Vernois, M.A., et al. 2001, \apj, 559, 296

\bibitem[Erlykin\&Wolfendale(1997)]{erlykin97}
Erlykin, A.D., \& Wolfendale, A.W. 1997, J. Phys.\ G23, 979

\bibitem[Fanselow et al.(1969)]{fanselow69}
Fanselow, J.L., Hartman, R.C., Hildebrand, R.H., \& Meyer, P.
1969 \apj 158 771

\bibitem[Gleeson \& Axford(1967)]{gleeson67}
Gleeson, L.J., \& Axford, W.I. 1967, \apj, 149, L115

\bibitem[Gleeson \& Axford(1968)]{gleeson68}
Gleeson, L.J., \& Axford, W.I. 1968, \apj, 154, 1011

\bibitem[Gleeson \& Urch(1973)]{gleeson73}
Gleeson, L.J., \& Urch, I.H. 1973, \ass, 25, 387

\bibitem[Golden et al.(1984)]{golden84}
Golden, R.L., Mauger, B.G., Badhwar, G.D., Daniel, R.R., 
Lacy, J.L., Stephens, S.A., \& Zipse, J.E. 1984, \apj, 287, 622

\bibitem[Golden et al.(1994)]{golden94}
Golden, R.L., et al. 1994, \apj, 436, 769

\bibitem[Green(2001)]{green01}
Green, D.A. 2001,
`A Catalogue of Galactic Supernova Remnants (Dec.~2001)',
Mullard Radio Astronomy Observatory, Cavendish Laboratory, Cambridge,
United Kingdom, \url{http://www.mrao.cam.ac.uk/surveys/snrs/}

\bibitem[Grimani et al.(2002)]{grimani02}
Grimani, C., et al. 2002, \aea, 392, 287

\bibitem[Guetta\&Amato(2003)]{guetta03}
Guetta, D. \& Amato, E.\ 2003, Astropart. Phys. 19, 403

\bibitem[Heber\&Potgieter(2000)]{heber00}
Heber, B. \& Potgieter, M.S. 2000, Adv.\ Space Res.\ vol.\ 26, no.\ 5, 839

\bibitem[Kobayashi et al.(2001)]{kobayashi01}
Kobayashi, T., et al. 2001,  \asr vol. 27, no. 4, 653

\bibitem[Ma\'{\i}z-Apell\'aniz(2001)]{maiz01}
Ma\'{\i}z-Apell\'aniz, J.\ 2001, ApJ 560, L83

\bibitem[Maurin et al.(2002)]{maurin02}
Maurin, D. et al.\ 2002, A\&A 394, 1039

\bibitem[Meegan \& Earl(1975)]{meegan75}
Meegan, C.A., \& Earl, J.A. 1975, \apj, 197, 219

\bibitem[M\"uller(2001)]{muller01}
M\"uller, D. 2001,  \asr vol. 27, no. 4, 659

\bibitem[Nishimura et al.(1980)]{nishimura80}
Nishimura, J. et al. 1980, \apj, 238, 394

\bibitem[Nishimura et al.(2000)]{nishimura00}
Nishimura, J. et al. 2000, \asr vol. 26, no. 11, 1827

\bibitem[Parker(1965)]{parker65}
Parker, E.N. 1965, \pss, 13, 9

\bibitem[Pohl\&Esposito(1998)]{pohl98}
Pohl, M. \& Esposito, J.A. 1998, ApJ 507, 327

\bibitem[Potgieter\&Ferreira(1999)]{potgieter99}
Potgieter, M.S. \& Ferreira, S.E.S. 1999, Proc.\ of the
26$^\mathrm{th}$ ICRC, SH 3.1.06

\bibitem[Potgieter\&Ferreira(2001)]{potgieter01}
Potgieter, M.S. \& Ferreira, S.E.S. 2001, Adv.\ Space Res.\ vol.\ 27,
no.\ 3, 481

\bibitem[Prince(1979)]{prince79}
Prince, T.A. 1979, \apj, 227, 676

\bibitem[Schlickeiser(2003)]{schli03}
Schlickeiser, R. 2002, \emph{Cosmic Ray Astrophysics}, Berlin,
Springer-Verlag

\bibitem[Stephens(2001)]{stephens01}
Stephens, S.A. 2001,  \asr vol. 27, no. 4, 687

\bibitem[Tang(1984)]{tang84}
Tang, K.-K. 1984, \apj, 278, 881

\bibitem[Thorsett et al.(2003)]{thorsett03}
Thorsett, R.A., et al.\ 2003, ApJ, 592, L71

\bibitem[Torii et al.(2000)]{torii00}
Torii, S., et al. 2000, \asr vol. 26, no. 11, 1867

\bibitem[Torii et al.(2001)]{torii01}
Torii, S., et al. 2001, \apj, 559, 973

\bibitem[Usoskin et al.(2003)]{usoskin03}
Usoskin, I.G., et al. 2003, Proc.\ of the 28$^\mathrm{th}$ ICRC,
SH 3.4, 4041

\bibitem[V\"olk(2003)]{voelk03}
V\"olk, H.J., 2003, private communication

\bibitem[Webber \& Lockwood(2001)]{webber01}
Webber, W.R. \& Lockwood, J.A. 2001, \jgr, vol.\ 106, no.\ A12, 29323

\end{thebibliography*}
%\end{thebibliography}

\clearpage

\begin{deluxetable}{lcccccc}
%\tabletypesize{\scriptsize}
\tablecaption{Direct Measurements of Cosmic Ray Electrons. \label{tbl-1}}
\tablewidth{0pt}
\tablehead{
\colhead{Measurement} & 
\colhead{Year} &
\colhead{Sun} &
\colhead{$\mathrm{e^-/e^+}$} &
\colhead{$E_\mathrm{min}$} &
\colhead{$E_\mathrm{max}$} &
\colhead{Ref.} \\
\colhead{} &
\colhead{} &
\colhead{pol.} &
\colhead{sep.} &
\colhead{(GeV)} &
\colhead{(GeV)} &
\colhead{} 
}
\startdata
Fanselow 1969    & 1965, 1966 & --  & Y &  0.07 &  11.0 &  1 \\
Nishimura 1980   & 1968--1975 & --,+& N & 30.0  & 1500  &  2 \\
Buffington 1975  & 1972, 1973 & +   & Y &  5.1  &  63.0 &  3 \\
Meegan 1975      & 1969, 1973 & +   & N &  6.4  & 114   &  4 \\
Prince 1979      &   1975     & +   & N & 10.2  & 202   &  5 \\
Golden 1984      &   1976     & +   & N &  3.45 &  91.7 &  6 \\
Tang 1984        &   1980     & +   & N &  4.89 & 200   &  7 \\
MASS89           &   1989     & --  & Y &  1.6  &  16.1 &  8 \\
MASS91           &   1991     & +   & Y &  7.5  &  46.9 &  9 \\
CAPRICE94        &   1994     & +   & Y &  0.54 &  34.3 & 10 (e), 11 (p) \\
HEAT94           &   1994     & +   & Y &  5.45 &  66.4 & 12 \\
HEAT95           &   1995     & +   & Y &  1.20 &  66.4 & 12 \\
Nishimura 2000   & 1996, 1998 & +   & N & 30.0  & 3000  & 13 \\
BETS97+98        & 1997, 1998 & +   & N & 13.9  & 112.6 & 14 \\
AMS-01           &   1998     & +   & Y &  0.15 &  35.7 & 15 (e), 16 (p) \\
\enddata
\tablecomments{Only experiments which published data tables are
 reported (the AMS Collaboration provided the numerical tables in
 electronic format).  Positive and negative Sun polarities refer to
 epochs when the magnetic field emerging from the North Pole of the
 Sun points outward and inward, respectively \citep{bieber99}.}
\tablerefs{
 (1) \citealt{fanselow69};   % 1965, 1966
 (2) \citealt{nishimura80};  % 1968--1975
 (3) \citealt{buffington75}; % 1972, 1973
 (4) \citealt{meegan75};     % 1973
 (5) \citealt{prince79};     % 1975
 (6) \citealt{golden84};     % 1976
 (7) \citealt{tang84};       % 1980
 (8) \citealt{golden94};     % 1989         MASS89
 (9) \citealt{grimani02};    % 1991         MASS91
(10) \citealt{boezio00};     % 1994         CAPRICE94 e+ e-
(11) \citealt{boezio99};     % 1994         CAPRICE94 p H
(12) \citealt{duvernois01};  % 1994, 1995   HEAT94, HEAT95
(13) \citealt{nishimura00};  % 1996, 1998
(14) \citealt{torii01};      % 1997, 1998   BETS97+98
(15) \citealt{ams00el}.      % 1998         AMS98 e+ e-
(16) \citealt{ams00pr2}.     % 1998         AMS98 protons
}
\end{deluxetable}

\end{document}